\documentclass[%
reprint,
nobibnotes,
superscriptaddress,
amsmath,amssymb,
aps,
floatfix,
]{revtex4-1}

\usepackage{siunitx}
\usepackage{bm}%
\usepackage{xcolor} 
\usepackage{graphicx}
\usepackage[utf8]{inputenc}
\usepackage{epstopdf}
\usepackage{afterpage}
\usepackage{setspace}
\usepackage{booktabs}
\usepackage{hyperref}
\usepackage[table]{xcolor}

\expandafter\ifx\csname package@font\endcsname\relax\else
 \expandafter\expandafter
 \expandafter\usepackage
 \expandafter\expandafter
 \expandafter{\csname package@font\endcsname}%
\fi
\begin{document}

\setstretch{1.0}
\title{Atomistic simulations of electrochemical systems with machine learning potentials}

\author{Marialore Sulpizi}
\email{marialore.sulpizi@rub.de}
\affiliation{Department of Physics and Astronomy, Ruhr-Universit\"at Bochum, 44780 Bochum, Germany}
\author{J\"{o}rg Behler}
\email{joerg.behler@rub.de}
\affiliation{Lehrstuhl f\"ur Theoretische Chemie II, Ruhr-Universit\"at Bochum, 44780 Bochum, Germany}
\affiliation{Research Center Chemical Sciences and Sustainability, Research Alliance Ruhr, 44780 Bochum, Germany}

\date{\today}

\begin{abstract}
Modern machine learning potentials (MLPs) enable large-scale atomistic simulations at unprecedented accuracy. The incorporation of explicit physical interactions such as long-range electrostatics has continuously broadened their applicability to increasingly complex systems in chemistry and materials science. Extending MLPs to electrochemistry and electrocatalysis, however, is still challenging as it requires the consideration of electrode potentials and electric fields in the atomic interactions. Such approaches have emerged only recently, and although the field is evolving rapidly, further conceptual advances are still needed for carrying out reliable simulations at finite temperature under applied potential. In this review, we provide a comprehensive overview about recent progress in the atomistic simulation of electrified interfaces using MLPs. Current developments fall into three different categories. The first comprises linear response models that describe polarisation, polarisability, and Born effective charges, extending conventional MLPs to finite electric fields. The second includes methods that directly learn a field-dependent potential energy surface. The third consists of a series of hybrid approaches combining classical models with machine learning. Although still under active development, these methods represent a major step toward more realistic simulations approaching the quality of electronic structure calculations. Finally, we discuss open challenges and possible future directions for advancing MLP-based simulations aiming for a bottom-up understanding of atomistic processes at electrified interfaces.
\end{abstract}

\maketitle

\section{Introduction}\label{introduction}

Advances in the electrochemical sciences are essential to address the technological and environmental challenges we are facing. There is an urgent need for a rapid electrification of our society away from fossil energy sources to reduce greenhouse gas emissions~\cite{P7384,P7432}. This need has, e.g., led to a paradigm change in mobility towards electric vehicles relying on modern batteries. Apart from transportation, advances in battery technology are equally important in many other fields, from mobile phones and laptops via short-term energy storage in home solar systems to emerging autonomous robots. 
Another route to electrochemical energy storage from renewable sources at large scale is the production of green hydrogen by water splitting~\cite{P7284,P7437}. Next to generating electricity in fuel cells, hydrogen offers new possibilities for the decarbonization of industrial processes by replacing fossil fuels as the primary energy source~\cite{P6716}. Regarding CO$_2$, carbon capture and conversion do not only offer a reduction of greenhouse gases in the atmosphere, but also open a path to alternative industrial feedstocks. Electrified synthesis routes and synthetic fuels~\cite{P7285} are just emerging with the ultimate goal of reaching a closed carbon cycle~\cite{P7438}. The search for the required materials does not only include the identification of improved battery electrodes~\cite{P7385} and more efficient catalysts for the electrification of chemical processes~\cite{P6694,P6717}, but also the development of more resistant materials to reduce the economic damage caused by corrosion~\cite{P7019}. Finally, electrochemical sensors enable completely new approaches, from monitoring water quality~\cite{P7436} to medical diagnostics~\cite{P7435}. 

Understanding electrochemistry and electrocatalysis at the atomic level and gaining insights into the underlying atomistic mechanisms is vital for further technological advances. Interfaces between electrolytes and solids play a central role in these processes, and due to the complexity of these systems, combining experimental data with information from computer simulations is essential for further progress~\cite{P6462}.

Electrochemical interfaces pose a significant challenge for atomistic simulations. Modeling a liquid electrolyte requires very large systems to yield reliable structural and dynamical properties based on converged statistical sampling. This does not only concern the configurations of the solvent molecules, but also a realistic representation of the concentrations and distributions of ions including their solvation spheres, which are essential to avoid finite-size effects and to gain insights into the properties of the electric double layer. Moreover, ``real electrodes'' are far from ideal single crystal surfaces. 
Instead, the structure of an electrode can depend on its history and kinetic factors and may exhibit a wide range of defects and even potential-induced surface reconstructions~\cite{P6045,P7439}, which need to be taken into account~\cite{P7345}. Further, electrocatalytic processes are not occurring at static interfaces, but are dynamic in nature, involving many different phenomena, like the diffusion of ions and reactants as well as metastable states and frustrated phase transitions at the catalyst surface~\cite{P7440}. Finally, another level of complexity is represented by the formation of amorphous solid-electrolyte interphases (SEI)~\cite{P7430}, for which little structural information is available. 
This high complexity of real electrochemical interfaces is a fundamental challenge requiring computer simulations of very large systems at extended timescales with the long-term goal to get as close as possible to operando conditions to extract essential information about these systems.

Apart from the development of relevant structural interface models, another challenge is the description of the atomic interactions. While electronic structure methods like density-functional theory (DFT)~\cite{P7442} are well-established in heterogeneous catalysis for gas-surface reactions, the sampling of liquid electrolytes at finite temperatures strongly increases the computational costs of DFT-based ab initio molecular dynamics (MD)~\cite{B0007} drastically limiting the number of solvent molecules in the simulations. Moreover, the reliability of DFT calculations depends on the employed exchange-correlation functional~\cite{P7441}, and often the particular choice is a compromise achieving a reasonable overall description of the electrolyte and the solid electrode. Finally, in spite of decades of research, the inclusion of explicit electrode potentials in DFT calculations is still an active field of research, which has led to a variety of different approaches, each with its own advantages and limitations~\cite{P7351,P7287}. 

Due to the high computational costs of electronic structure calculations, extending the time and length scales of atomistic simulations beyond the realm of ab initio MD traditionally required the use of more approximate methods to compute the energies and forces, which unavoidably results in a decrease of accuracy. A balanced combination of empirical potentials and classical force fields describing the electrode and electrolyte subsystems is often difficult. Another severe limitation of many simple water models is their inability to describe the making and breaking of bonds~\cite{P2803}. This lack of reactivity prevents to account for the vital role of water dissociation, recombination and proton transfer at aqueous electrified interfaces. The consideration of protons in full generality requires the use of explicit, i.e., fully atomistic and reactive solvents, although also implicit solvents have received considerable attention as a pragmatic solution in some cases~\cite{P6461}. Further challenges are the inclusion of nuclear quantum effects~\cite{P6303}, pH and the very different time scales involved in electrochemical processes~\cite{P7346}, which all require approaches beyond standard classical MD simulations.

In recent years, machine learning (ML) techniques have emerged as powerful new tools in many fields of science. One of the early and particularly fruitful applications of ML in chemistry and materials science is the representation of the potential energy surface (PES), which provides the atomic interactions as a function of the structure of a system. The resulting machine learning potentials (MLP) now enable to compute the energies and forces with the quality of quantum mechanical methods like DFT at drastically reduced costs~\cite{P4885,P5673,P5793,P6102,P6112,P6121,P6631}. Consequently, MLPs now allow to transfer the accuracy of ab initio MD to more complex systems at nanosecond timescales, which greatly facilitates the sampling of explicit solvents in interfacial systems~\cite{P7208}. Moreover, MLPs are reactive such that key processes like proton transfer reactions become accessible~\cite{P5599} lifting critical restrictions and tradeoffs of conventional force field-based simulations. Still, in spite of numerous application of MLPs to solid-liquid interfaces published to date~\cite{P6721}, the simulation of electrified interfaces using MLPs is still a young and emerging field. 

In this review we provide an overview of the current state-of-the-art in the simulation of electrochemical systems employing MLPs, which has received increasing attention in recent years~\cite{P6007,P6361,P6468,P6905,P7278,P7297,P7298,P7307,P7389}. While the application of ML itself in electrochemistry is much broader~\cite{P6768,P7321}, here we will focus on atomistic simulations with the aim to gain bottom-up atomic-level insights into the underlying processes. We do not cover other applications of ML in electrochemistry, such as materials design, the exploration of chemical space by high-throughput screening in electrocatalysis~\cite{P7443}, the analysis of experimental data~\cite{P6475} or self-driving labs~\cite{P7491}.

In the next Section, we start with a short overview of machine learning potentials, restricting ourselves to some selected examples in this rapidly evolving field. We then give a brief summary of application of MLPs to systems and system classes like water, electrolytes and solid-liquid interfaces, which have been studied for many years, mainly in the absence of electric fields and electrode potentials. In Section~\ref{sec:electrified-interfaces} we continue with a wrap-up of the ``conventional'' simulation of electrified interfaces using electronic structure methods and classical models. Many important concepts have been developed over the years, which are now in the process of being transferred to the field of MLPs. In particular, DFT calculations of electrified interfaces will remain important as source for the training data needed in the construction of MLPs.
In Section~\ref{sec:elecMLP} we then review different approaches that have been proposed for the simulation of electrochemical systems using MLPs. After discussing the case of redox potentials, we provide an overview of the three main strategies which currently receive most attention. These are linear response methods, the direct learning of field-dependent MLPs, and hybrid methods combining machine learning with classical models. In Section~\ref{sec:discussion} we conclude with a discussion of the remaining challenges and an outlook to possible future developments in the field.

\section{Machine Learning Potentials}\label{sec:MLPs}

\subsection{A concise overview of methods}

The advent of MLPs has revolutionized the field of atomistic simulations. Once trained to energies and forces obtained from electronic structure calculations in a supervised learning process, they allow to significantly extend both, system size and simulation time, while essentially maintaining the accuracy of the underlying reference method~\cite{P4885,P5673,P6112,P6102,P6121}. These capabilities, along with many successful applications, have stimulated the development of a large number of different approaches, which share the same overall scope but are based on very different ideas and algorithms. Some selected examples for different types of MLPs are given in Table~\ref{tab:MLP}. Keeping track of this large variety of methods is very challenging, and assessing their applicability and limitations may be facilitated by a classification of MLPs into different generations~\cite{P6018}. There are many possible criteria for such a classification, and one way is based on the physical interactions that are included and, as a result, the systems that can be studied.

First-generation (1G) MLPs have been introduced about three decades ago by Doren and coworkers~\cite{P0316}. They relied exclusively on neural networks, and the description of chemistry at interfaces has been a major motivation for the development of these early MLPs~\cite{P0316,P0421,P1388}. However, a main limitation of 1G-MLPs is their restriction to very small systems containing only a few mobile atoms and their associated degrees of freedom. A serious conceptual problem at that time preventing a wider use of these methods has been the lack of generally applicable representations of the atomic positions. Coordinates suitable as ML input features must ensure the invariances of the PES with respect to rotation and translation of the system and permutations in the order of chemically equivalent atoms~\cite{P2882}. A review of early 1G-MLPs constructed for low-dimensional PESs can be found in Ref.~\cite{P3033}.

MLPs became a practical tool for simulations of condensed systems containing thousands of atoms with the introduction of the second generation (2G) exploiting the decay of the atomic interactions with increasing distance. The resulting assumption of short-ranged atomic interactions is a surprisingly good approximation for many systems. Starting with high-dimensional neural network potentials (2G-HDNNP) proposed in 2007~\cite{P1174}, the total energy of 2G-MLPs is constructed as a sum of environment-dependent atomic energies, which are added to yield the total energy of the system. The decomposition of the system's energy is done automatically by the MLP in the training process using reference total energies. Due to the locality of the atomic energies, 2G methods are linear scaling in the number of atoms enabling efficient large-scale simulations for systems much larger than those used in training. 

There are three families of 2G-MLPs, which differ in the way the local atomic energies are constructed. The first one combines predefined geometric features, like atom-centered symmetry functions (ACSFs)~\cite{P2882} in case of 2G-HDNNPs~\cite{P1174} or smooth overlap of atomic positions (SOAP)~\cite{P3885} in case of Gaussian approximation potentials (GAP)~\cite{P2630}, with non-linear ML methods like neural networks or Gaussian kernels, and many other descriptor-based approaches have been proposed~\cite{P5596,P6608,P5792,P4945,P7380}. Other methods like atomic cluster expansion (ACE)~\cite{P5794}, moment tensor potentials (MTP)~\cite{P4862} or spectral neighbor analysis potentials (SNAP)~\cite{P4644} construct the atomic energies using linear regression of non-linear functions capturing the atomic environments. All these methods typically define a fixed-size local environment employing a cutoff radius between 5 and 10~\AA{}. A cutoff function ensures a smooth decay of the atomic energies and forces to zero at the cutoff distance to avoid discontinuities in these properties, e.g., during MD simulations. The required cutoff can be rigorously determined based on the interactions in the system~\cite{P6312,P4958}.
In the third family, starting with the introduction of deep tensor neural networks (DTNN) in 2017~\cite{P4937}, message passing (MP)-based MLPs~\cite{P5368} employ graph neural networks to obtain an extended but still finite local environment by passing information from atom to atom over several cutoff radii through a small number of message passing steps~\cite{P5366,P5817,P6787}. Hence, sometimes these methods are called ``semilocal''. More recent methods incorporate information-rich equivariant features~\cite{P6017,P6572,P6569}. Overall, a very large number of 2G-MLPs is now available employing many different ML techniques, and due to their relatively low computational costs, they have now become the workhorse methods in atomistic simulations.

By construction, a limitation of local 2G-MLPs is their finite range, because interactions extending beyond the local environment, such as electrostatics and dispersion, may be important for some systems~\cite{P5975}. Such longer-ranged interactions are explicitly included in third-generation (3G) MLPs. 
In this generation, charges~\cite{P2962,P5313,P6200} and multipoles~\cite{P2041} for electrostatics or Hirshfeld volumes for the computation of dispersion interactions~\cite{P6580,P7126} are learned as local atomic properties. These are then used in combination with explicit physical functional forms like Coulomb's law to yield long-range interactions. Moreover, the use of latent properties, which provide a higher degree of flexibility without the need for reference charges from electronic structure calculations, has been proposed~\cite{P7233}. Next to 3G-MLPs making use of atomic environments up to a cutoff radius to represent the atomic properties, also message passing MLPs have been extended to learn atomic charges for computing electrostatic interactions~\cite{P5577,P6037,P6612}. 

In spite of including long-range interactions, 3G-MLPs still have limitations arising from the locality of the underlying atomic properties. Most notably, they are unable to describe charge transfer over distances larger than the local atomic environments or the receptive fields of message passing neural networks. Such long-range charge transfer is omnipresent in many systems and naturally included in self-consistent electronic structure calculations. It is important to note that the resulting non-local interactions, which also affect short-range bonding due to the redistribution of electron density in the system, are different from the long-range interactions included in 3G-MLPs, as the latter cannot account for these global phenomena. 

Non-local interactions arising from global charge redistributions are important for many types of systems, from (de)protonation reactions of large organic molecules, possibly changing the charge density in the entire molecule, to defects or doping in materials like oxide surfaces~\cite{P5977,P6018}. Consequently, a fourth-generation (4G) of MLPs making use of global information is required for such systems. Starting with the introduction of CENT in 2015~\cite{P4419}, a series of 4G methods has been introduced~\cite{P5932,P6532,P6666,P6122}, which include a charge equilibration (QEq)~\cite{P1448} step to distribute the charge in the system based on information about its global structure. Other 4G-MLPs employ alternative techniques to describe non-local charge transfer, e.g. a self-consistent determination of the atomic charges in BpopNN~\cite{P5859}, attention mechanisms in SpookyNet~\cite{P6056}, the use of learned Fukui functions in MACE-polar-1~\cite{P7251} or the construction of global equivariant descriptors in LOREM~\cite{P7281}.

On top of these ongoing developments aiming to incorporate more and more physical concepts directly into MLP architectures, two general research directions can be identified. One direction targets for the accurate and very efficient description of the PES for a particular system of interest using slim and fast to evaluate MLPs enabling simulations of tens of thousands of atoms on cheap and often CPU-based machines. The goal of the second direction is to derive general-purpose MLPs often termed ``foundation models'', which are as transferable as possible across chemical space with the ultimate goal to reduce the effort in constructing MLPs for new systems~\cite{P6669,P7098,P7238,P7251,P7517,P6232}. This comes at the cost of more complex architectures with often very large numbers of parameters, making the direct use of these models computationally more demanding, often requiring powerful GPU-based machines. These costs and also the lower accuracy can be improved in system-specific fine-tuning and destillation of more efficient derived models.

It is beyond the scope of this review to provide an exhaustive overview of the huge number MLPs that has been published to date, and more details can be found in several dedicated reviews covering a large variety of MLP methods~\cite{P4885,P5673,P5793,P6102,P6112,P6121,P6131}.

\begin{table*}[!htbp]
\caption{Evolution of MLPs illustrated for selected examples. The background color represents the underlying physical model (blue: second-generation using local atomic energies; yellow: third-generation including long-range electrostatics; green: consideration of global non-local charge transfer; grey: others). Methods written in italics contain message passing, methods in bold represent foundation models.}\label{tab:MLP}
\footnotesize
\begin{tabular}{ccccccccccccccccccccccc} 
  \rotatebox{90}{1995} & ...    & 
  \rotatebox{90}{2007} & \rotatebox{90}{2008} & \rotatebox{90}{2009} & \rotatebox{90}{2010} & 
  \rotatebox{90}{2011} & \rotatebox{90}{2012} & \rotatebox{90}{2013} & \rotatebox{90}{2014} & 
  \rotatebox{90}{2015} & \rotatebox{90}{2016} & \rotatebox{90}{2017} & \rotatebox{90}{2018} & 
  \rotatebox{90}{2019} & \rotatebox{90}{2020} & \rotatebox{90}{2021} & \rotatebox{90}{2022} & 
  \rotatebox{90}{2023} & \rotatebox{90}{2024} & \rotatebox{90}{2025} & \rotatebox{90}{2026} &\\ 
 \multicolumn{23}{l}{\cellcolor{gray!20} Early first-generation MLPs~\cite{P0316}}  \\
 & & \multicolumn{21}{l}{\cellcolor{blue!20} 2G-HDNNP~\cite{P1174}}  \\
 & & & & & \multicolumn{18}{l}{\cellcolor{blue!20} GAP~\cite{P2630}}  \\
 & & & & & & & \multicolumn{16}{l}{\cellcolor{yellow!40} 3G-HDNNP~\cite{P2962}}  \\
 & & & & &     & & & & & \multicolumn{13}{l}{\cellcolor{blue!20} SNAP~\cite{P4644}}\\
 & & & & &     & & & & & \multicolumn{13}{l}{\cellcolor{green!20} CENT~\cite{P4419}}\\
 & & & & &     & & & & &    &  \multicolumn{12}{l}{\cellcolor{blue!20} MTP~\cite{P4862}}\\
 & & & & &     & & & & &    & & \multicolumn{11}{l}{\cellcolor{blue!20} DTNN~\cite{P4937}}\\
 & & & & &     & & & & &    & & \multicolumn{11}{l}{\cellcolor{blue!20} ANI-1~\cite{P4945}}\\      
 & & & & &     & & & & &    & & & \multicolumn{10}{l}{\cellcolor{yellow!40} Tensormol~\cite{P5313}}\\
 & & & & &     & & & & &    & & & \multicolumn{10}{l}{\cellcolor{blue!20} DeepMD~\cite{P5596}}\\
 & & & & &     & & & & &    & & & \multicolumn{10}{l}{\cellcolor{blue!20} \emph{SchNet}~\cite{P5366}}\\
 & & & & &     & & & & &    & & & & \multicolumn{9}{l}{\cellcolor{yellow!40} \emph{PhysNet}~\cite{P5577}}\\
 & & & & &     & & & & &    & & & & \multicolumn{9}{l}{\cellcolor{blue!20} ACE~\cite{P5794}}\\
 & & & & &     & & & & &    & & & & \multicolumn{9}{l}{\cellcolor{blue!20} \emph{AIMNet}~\cite{P5817}}\\
 & & & & &     & & & & &    & & & & \multicolumn{9}{l}{\cellcolor{blue!20} EANN~\cite{P5792}}\\
 & & & & &     & & & & &    & & & & & \multicolumn{8}{l}{\cellcolor{green!20} BpopNN~\cite{P5859}}\\
 & & & & &     & & & & &    & & & & & \multicolumn{8}{l}{\cellcolor{blue!20} GMNN~\cite{P6608}}\\
 & & & & &     & & & & &    & & & & &   &  \multicolumn{7}{l}{\cellcolor{green!20} 4G-HDNNP~\cite{P5932}}\\
 & & & & &     & & & & &    & & & & &   &  \multicolumn{7}{l}{\cellcolor{blue!20} NEP~\cite{P7380}}\\    
 & & & & &     & & & & &    & & & & &   &  \multicolumn{7}{l}{\cellcolor{green!20} \emph{SpookyNet}~\cite{P6056}}\\
 & & & & &     & & & & &    & & & & &   &  \multicolumn{7}{l}{\cellcolor{blue!20} \emph{PAINN}~\cite{P6026}}\\
 & & & & &     & & & & &    & & & & &   & & \multicolumn{6}{l}{\cellcolor{blue!20} M3GNet~\cite{P6232}}\\  
 & & & & &     & & & & &    & & & & &   & & \multicolumn{6}{l}{\cellcolor{green!20} QRNN~\cite{P6122}}\\
 & & & & &     & & & & &    & & & & &   & & \multicolumn{6}{l}{\cellcolor{yellow!40} DPLR~\cite{P6200}}\\
 & & & & &     & & & & &    & & & & &   & & \multicolumn{6}{l}{\cellcolor{blue!20} \emph{NequIP}~\cite{P6017}}\\
 & & & & &     & & & & &    & & & & &   & & \multicolumn{6}{l}{\cellcolor{blue!20} \emph{MACE}~\cite{P6572}}\\
 & & & & &     & & & & &    & & & & &   & & \multicolumn{6}{l}{\cellcolor{green!20} kQEq~\cite{P6666}}\\
 & & & & &     & & & & &    & & & & &   & & & \multicolumn{5}{l}{\cellcolor{blue!20} Allegro~\cite{P6569}}\\
 & & & & &     & & & & &    & & & & &   & & & \multicolumn{5}{l}{\cellcolor{green!20} PANNA2~\cite{P6532}}\\
 & & & & &     & & & & &    & & & & &   & & & \multicolumn{5}{l}{\cellcolor{blue!20} \emph{\textbf{CHGNet}}~\cite{P7098}}\\
 & & & & &     & & & & &    & & & & &   & & & & \multicolumn{4}{l}{\cellcolor{blue!20} \emph{GRACE}~\cite{P6787}}\\
 & & & & &     & & & & &    & & & & &   & & & & & \multicolumn{3}{l}{\cellcolor{yellow!40} \emph{\textbf{AIMNet2}}~\cite{P6612}}\\
 & & & & &     & & & & &    & & & & &   & & & & & \multicolumn{3}{l}{\cellcolor{blue!20} \emph{\textbf{UMA}}~\cite{P7238}}\\
 & & & & &     & & & & &    & & & & &   & & & & & \multicolumn{3}{l}{\cellcolor{blue!20} \textbf{Allegro-FM}~\cite{P7137}}\\
 & & & & &     & & & & &    & & & & &   & & & & & \multicolumn{3}{l}{\cellcolor{blue!20} \emph{\textbf{MACE-MP-0}}~\cite{P6669}}\\
 & & & & &     & & & & &    & & & & &   & & & & &  &\multicolumn{2}{l}{\cellcolor{blue!20} \emph{\textbf{GRACE-FM}}~\cite{P7517}}\\
 & & & & &     & & & & &    & & & & &   & & & & &  &\multicolumn{2}{l}{\cellcolor{green!20} \emph{\textbf{MACE-POLAR-1}}~\cite{P7251}}\\
\end{tabular}
\end{table*}

\subsection{Machine Learning Potentials for Solid-Liquid Interfaces}

\begin{figure}
  \centering
    \includegraphics[width=0.45\textwidth]{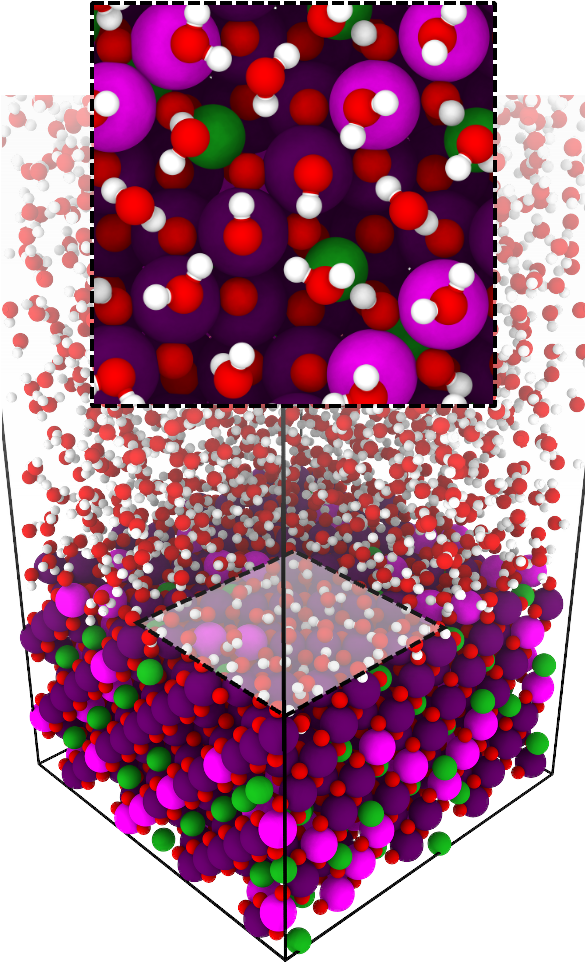}
    \caption{HDNNP simulation of water dissociation at the (001) surface of the cathode material LiMn$_2$O$_4$ revealing the mechanism underlying its surface degradation. After the dissociation of a water molecule at the Lewis acidic Mn$^{4+}$ site acting as proton source (dark violet), the proton is transferred to the Mn$^{3+}$ site (light magenta), where the acid-induced disproportionation of Mn$^{3+}$ into Mn$^{4+}$ and water-soluable Mn$^{2+}$ according to the Hunter reaction takes place. Li atoms are shown in green, oxygen in red and hydrogen in white~\cite{P7449}.
    }\label{fig:LiMn2O4}
\end{figure}

Although the development of MLPs is still a very active field of research, modern MLPs have now reached a level of maturity that allows to perform large-scale simulations of a variety of systems. Here, we provide some selected MLP applications that are closely related to electrochemistry, i.e., liquid water, electrolytes and aqueous solid-liquid interfaces. The examples discussed in this section do not include electrode potentials or electric fields (see Section~\ref{sec:elecMLP}), and a more detailed discussion of MLP-driven simulations in the absence of electrode potentials and electric fields can be found in several recent reviews~\cite{P6721,P7208,P6058}. 

Due to its fundamental importance, water has received a lot of attention in atomistic simulations~\cite{P2803}, and its properties governed by hydrogen bonding make it a challenging benchmark system for the development of interatomic potentials. A lot of effort has been put in the investigation of water using ab initio MD, but a key question has remained unanswered for a long time: How good is the DFT description of water, and to what extent does it depend on the chosen exchange correlation functional~\cite{P6212}? MLPs now can extend the time and length scales of simulations of water without loss in accuracy allowing to address this question for properties that are too demanding to compute with on-the-fly ab initio MD.
For instance, first MLPs published for bulk liquid water and ice polymorphs studied the role of dispersion corrections in combination with different GGA functionals~\cite{P4556}. Several other studies have addressed the role of the functional~\cite{P5479,P7334,P7373}, the damping function in dispersion interactions~\cite{P7126} and the impact of nuclear quantum effects~\cite{P4586,P7349,P7214}. A wide range of properties of water has now been studied with different types of neural network potentials~\cite{P5639,P5711,P6854,P6150,P6151,P6195,P6196}, mainly relying on HDNNPs~\cite{P1174} and DeepMD~\cite{P5596}, and even water models approaching the coupled cluster level of accuracy have now become available~\cite{P6360,P6362,P6321}.

Next to pure water, aqueous electrolytes are of high relevance for studying electrochemical interfaces~\cite{P6007}. Detailed HDNNP studies have been reported for the structure and proton transfer mechanisms in aqueous NaOH solutions over a wide concentration range~\cite{P4670,P4895}, and a novel proton transfer-driven water exchange mechanism in the Na$^+$ solvation shell could be identified~\cite{P5126}. The proton transfer rate in this process is significantly accelerated when nuclear quantum effects are explicitly considered~\cite{P5631}. Maybe the most prototypical system for electrolytes is NaCl, and its dissolution process has been extensively studied~\cite{P7494} along with properties of its aqueous solution~\cite{P6292,P7496,P7501,P7502}. A MLP trained to MP2 data has shown that contact-ion pairs and solvent-shared ion pairs possess nearly identical stabilities~\cite{P7220}.

Further systems that have been studied by MLPs are solutions of CaCl$_2$~\cite{P7069}, KCl~\cite{P6292}, KOH~\cite{P7507}, CsI~\cite{P7502} and NaBr~\cite{P6292} as well as the Zn$^{2+}$~\cite{P5615}, fluoride~\cite{P6058} and sulfate ions~\cite{P6058}. Finally, Li cation dimer formation has been investigated using NequIP~\cite{P7498}, while charge recursive neural networks have addressed the properties of liquid electrolytes in Li ion batteries~\cite{P6338}. 
Simulations have also been extended to electrolyte-air interfaces, and a combination with vibrational sum-frequency generation addressing the distribution of ions close to the interface provided evidence for an ion-depleted outer layer above an ion-enriched subsurface region~\cite{P7495}. 
Very recently, studies have also been reported for more complex molecular ions in aqueous solution, such as 2'-deoxyadenosine 5'-monophosphate (dAMP)~\cite{P7526}.

Of particular relevance for electrochemistry are solid-liquid interfaces, and many metal and oxide surfaces in contact with water have been investigated in recent years. In 2016, the interfaces between water and several low-index surfaces of copper have been studied~\cite{P4859}, and this work has later been extended to unravel the role of surface vacancies and adatoms on the properties of interfacial water~\cite{P4886}.
2G-HDNNPs have also been used to systematically understand the mechanisms of water dissociation and proton transfer at non-polar ZnO surfaces. It has been found that at ZnO(10$\bar{1}$0) there are two main mechanisms: a slow surface proton transfer between a dissociating water molecule and a surface oxygen atom, and a fast adlayer proton transfer between neighboring water molecules and hydroxide groups~\cite{P4988}. The transfer of protons has been found to be governed by the surrounding hydrogen-bond network in a presolvation-guided mechanism, similar to the proton transfer processes observed in NaOH solutions~\cite{P5126,P5601}. The dependence of proton transfer pathways on the specific ZnO surface geometry can lead to one-dimensional or two-dimensional proton transfer networks for different interfaces~\cite{P5599}. Computed vibrational frequencies can guide the characterization of the hydrogen bond patterns of the aqueous species close to the surface~\cite{P5600}. 

Another system, which has attracted a lot of attention, is water at TiO$_2$. In 2020, a DeepMD study of the anatase TiO$_2$–water interface addressed the dynamical equilibrium between molecular and dissociative adsorption, with dissociation occurring via solvent-assisted concerted proton transfer~\cite{P6271}, in line with the observations made for ZnO. Later work included the rutile polymorph of the same material~\cite{P6058,P6473} finding a higher degree of dissociation with implications for the activity of different surfaces in photocatalysis. In 2023 HDNNPs have been used to perform MD simulations of water at seven pristine TiO$_2$ surfaces of the anatase and rutile polymorphs systematically identifying a series of different proton transfer mechanisms~\cite{P7492}. 
Moreover, the TiO$_2$-water interface has been used as a benchmark system to establish an automated, active-learning framework for state-point-specific MLPs for complex aqueous systems~\cite{P6058}.

A broadly applicable MLP using the PaiNN architecture~\cite{P6026} has been developed for silica-water interfaces based on active learning strategies employing cluster structures~\cite{P6956}. Another study with the ChIMES MLP unraveled the water structure and dynamics at silica interfaces~\cite{P7506}.
A recent work using DeepMD demonstrates that nanoconfinement in a silica slit pore modifies the acidity of acetic acid depending on its spatial location~\cite{P7510}. 
Further HDNNP work addressed iron oxides like hematite and magnetite~\cite{P6661,P7119}, and the spinel Co$_3$O$_4$~\cite{P7181}. Moreover, several studies of calcium silicates have been carried out using different types of MLPs~\cite{P7237,P6582,P6752,P7114,P7152}. 

HDNNPs have also be used to study more complex oxides, like the interaction of water with several surfaces of LiMn$_2$O$_4$, a lithium intercalation compound used as cathode material in lithium ion batteries. Complex surface reactions and interface-dependent distributions of Mn$^{3+}$ and Mn$^{4+}$ ions could be identified in this material with interesting implications for possible catalytic applications~\cite{P6141}, since modifying the Li$^{+}$ content offers a route to tuning the surface electronic structure. Recently, it could be shown that the high Lewis acidity of Mn$^{4+}$ sites at the LiMn$_2$O$_4$(001) surface leads to water dissociation, with the emerging protons migrating to neighboring Mn$^{3+}$ sites as a first step of the Hunter reaction (s. Fig.~\ref{fig:LiMn2O4}), which ultimately leads to the acid-induced degradation of the material limiting its lifetime in cathode applications~\cite{P7449}.

Of particular interest in the context of electrochemistry are MD simulations of water at platinum surfaces. Investigated structural and energetic properties of liquid water and hydroxyl layers at Pt(111) show that the interface becomes increasingly hydrophobic at high hydroxyl coverages due to the reduction in hydrogen bonding between adsorbed water molecules and hydroxyl species as the surface becomes saturated~\cite{P6469}.
Other mechanistic insights have been gained for example regarding the influence of melamine adsorption on the oxygen reduction reaction at Pt(111) ~\cite{P7479}, and also the hydration free energies of oxygenated species at water–platinum interfaces have been determined using MLPs~\cite{P7476}.

Overall, it has been demonstrated that MLPs are a reliable tool to carry out first principles-quality simulations of a wide range of interface systems~\cite{P6859,P7020}. With the advent of foundation models such as MACE-MP-0, which has been employed in proof-of-principle studies to first solid-liquid interfaces and aqueous systems~\cite{P6669}, CHGNet~\cite{P7098}, Allegro-FM~\cite{P7137}, GRACE-FM~\cite{P7517} and many others, the number of MLP-driven simulations can be expected to grow substantially in the years to come~\cite{P7067}, although at present fine-tuning of these models is still required to reach the accuracy of DFT and of customized MLPs that are directly parameterized to represent specific systems.

\section{Electrified Interfaces}\label{sec:electrified-interfaces}
\subsection{Simulation Approaches}

Electrified interfaces have attracted considerable interest in atomistic simulations, with the development of computational methods accelerating over the past decades~\cite{P6315,P6317,P6318,P7286,P7287,P7288}. These advances have been driven by the need to deal with the intrinsically multiscale problems present in electrochemical systems. Still, in spite of the mature state reached by electronic structure methods like DFT for many applications, electrochemical interfaces pose additional specific challenges. Consequently, extensions have been required to study systems under constant charge or constant potential conditions~\cite{P7351}. A particular new focus has been on the role of charge transfer~\cite{P7318} including proton-coupled electron transfer~\cite{P7319} for interfacial reactivity, while connecting microscopic processes to experimentally relevant observables. 
With phenomena spanning from atomistic scales and picosecond dynamics of chemical reactions via nanosecond relaxations of the electrical double layer up to macroscopic scales and year-long timescales relevant to device operation and lifetime, also more approximate models of electrode processes have been developed to bridge these disparate scales. Still, addressing the complex, nonlinear interactions between them remains a major challenge.

\subsection{Density Functional Theory-based Methods}

We start with discussing the description of electrified interfaces by DFT, which as the by far most widely applied electronic structure method is not only important for direct studies of electrochemical systems in ab initio MD but also for the generation of MLP training data.

Simulations of electrified systems at the atomic level can be carried out using two main approaches: ``constant charge'' and ``constant potential''. 
In the constant-charge setup, the total charge of the system (or equivalently the total number of electrons) is predefined and fixed throughout the simulation and  the electrode potential fluctuates in response to changes in the surrounding environment. 
In the constant-potential approach, instead, the charges on the electrode can dynamically  fluctuate in response to the local electrostatic environment, while maintaining a prescribed electrode potential. This enables constant potential simulations to capture the electronic response of conductive electrodes to interfacial phenomena, such as, e.g., ion adsorption and rearrangements in the electrolyte, but also charge transfer between electrolyte and electrode.

There has been a long-standing discussion in the literature on the use of constant potential or constant charge approaches for the simulations of electrochemical reactions, but recent literature suggests some convergence. Using proton adsorption at metallic electrodes as a prototypical electrochemical problem the fundamental characteristics and practical applicability of both computational approaches have been compared~\cite{P7343}. The authors showed that both methods converge to identical differential free-energy changes in the thermodynamic limit and are thus both appropriate for equilibrium thermodynamic properties, where the two ensembles become equivalent~\cite{P7343}. They further discussed the improved interpretability and size convergence offered by the grand-canonical approach, while pointing out its increased sensitivity to specific modelling parameters. 

Santos and Schmickler have discussed the use of constant charge versus constant potential approaches in connection with the timescales of different electrochemical processes~\cite{P7346}. They suggest that electrochemical reactions are much faster than double-layer equilibration, therefore DFT and molecular dynamics simulations should generally be run at constant charge, not constant potential, when studying reaction kinetics. According to their observation, experimental potentiostats respond far too slowly to maintain a truly constant potential during ultrafast atomic-scale reactions.

In the next sections, we briefly review the most common constant change and constant potential approaches in DFT as a starting point for our later discussion of generalizing MLPs to electrified interfaces.

\subsubsection{Constant charge} 

Within the constant charge approach the electronic and ionic degrees of freedom are both treated canonically. The ionic coordinates are sampled thermally at fixed particle number, while the electronic charge degrees of freedom are constrained to satisfy the imposed total charge. This method is the easiest to implement in electronic structure codes and as been at the foundation of the very popular Computational Hydrogen Electrode (CHE) introduced by N{\o}rskov and coworkers in 2004~\cite{P3685}.
The key idea of this approach is to bypass the difficulty of simulating solvated protons $H^{+}$ and electrons $e^{-}$ by relating them to a simple, standard reference: hydrogen gas formation (\(1/2H_{2}(g)\rightleftharpoons H^{+}+e^{-}\)).
This has enabled the straightforward determination of reaction free energy diagrams at different potentials, and it has helped to identify potential-determining steps and optimal catalysts (e.g., explaining volcano plots)~\cite{P7314}.
The simulations are performed in the absence of an explicit bias and the reaction free energy can then be corrected for potential and pH contributions.

A first approach to explicitly simulate a net charge on the electrode was to vary the number of H atoms contained in the simulation box, resulting in the formation of H$^+$ ions in the electrolyte and negative charge on the metal slab in an overall neutral system \cite{P2844} due to Fermi level alignment.
This approach has, for example, been used to calculate the interfacial capacitance of the Pt(111)/water interface~\cite{P7488}. The concept can be generalized beyond proton excess by introducing an imbalance in the number of cations and anions in the electrolyte. In this ion-imbalance method, the electrode potential is controlled indirectly by varying the ionic composition of the solution, while the electrode acquires a compensating charge in response, such that the overall simulation cell remains charge neutral. An excess of anions in the solution compensates a positive charge on the metal in an overall neutral system, whereas an excess of cations can compensate a negatively charged electrode~\cite{P5588,P7465,P7487}. Across these studies, the ion-imbalance method evolved into a practical approach for simulating electrified Pt–water interfaces under explicit bias, enabling direct simulation of the double-layer structure, including water structure, hydrogen adsorption, wettability, and the role of surface morphology.

A key shortcoming of the constant-charge approach for electrochemical reactions is that the total system charge is fixed throughout the simulation, but in the absence of an electron reservoir the surface charge can still change in response to chemical processes involving electron transfer. This can be problematic for electrochemical reactions involving electron or proton–electron transfer, since due to the necessarily finite system size
significant changes in the electrode charge are unavoidable even if only one or a few electron transfer processes take place, which can completely alter the simulation conditions in small systems.  

\subsubsection {Constant potential}

The general challenge of constant potential methods in DFT is that fluctuating charges on the electrode must be realized under the constraint of an overall neutral system, which is mandatory in periodic DFT calculations. The available constant potential methods differ in how this neutrality is achieved.

An early first-principles formulation of a constant-potential description was introduced by Lozovoi, Alavi, Kohanoff and Lynden-Bell in 2001, who developed a periodic DFT scheme for charged slabs at constant electron chemical potential~\cite{P2880}. By defining the chemical potential of the electrode relative to a reference plane at a given distance in vacuum and introducing a compensating homogeneous background charge to preserve the neutrality of the periodically repeated simulation cell, their approach allowed the potential of the slab to be controlled while its total electronic charge could adjust accordingly. This provided an early grand-canonical formulation of an electrified surface in vacuum and established the fundamental distinction between simulations performed at fixed charge and those performed at fixed electronic chemical potential. The method has been employed to study charge-induced reconstruction of Pt(110) and Au(110) surfaces~\cite{P3664}.

A constant-potential periodic DFT study of a charged Pd(111) slab with compensating background charge covered by an explicit, thin ice-like water bilayer has been reported by Filhol and Neurock in 2006~\cite{P2878}. The method has also been applied to water at Cu(111) and methanol reactivity at Pt(111)~\cite{P2877}. In this work, the electrode potential is not imposed directly but tuned by changing the net charge on the metal surface through adding or removing electrons in the slab. A double reference method has been introduced to determine the free energies as a function of bias voltage by correcting for the interaction between the charged slab and the background charge and by referencing to the vacuum electrostatic potential.  

In 2006, Otani and Sugino developed the effective screening medium (ESM) method, which provided a description of the electrostatic boundary conditions of charged interfaces by modifying the pseudopotential plane-wave Poisson solver using a Green's function technique to lift periodicity of the system perpendicular to the electrode surface~\cite{P2876}. 
These developments led to the formulation of first-principles molecular dynamics at constant electrode potential~\cite{P2874}, and were used to investigate biased electrode/solution interfaces, including water/Pt systems~\cite{P2872}.
In 2012, the ESM-based MD has been further developed by Bonnet et al.~\cite{P3666} to treat the charges as additional dynamical variables that continuously adjust in response to the instantaneous configuration of the electrolyte exchanging with a reservoir via a Nos{\'e}–Hoover algorithm.

Building on joint DFT (JDFT) introduced by Letchworth-Weaver and Arias for efficient simulations of electrodes in contact with a continuum solvent~\cite{P3616}, in 2017 Sundararaman et al.~\cite{P7344} generalized and formalized this idea within a rigorous grand-canonical DFT (GC-DFT) framework, developing a method for minimizing the grand free-energy functional at fixed electron chemical potential.

In 2018, Melander et al. formulated electrochemical interfaces within a grand-canonical DFT framework, in which the electron and ion chemical potentials are fixed~\cite{P7477}. The ionic species are introduced through a coarse-grained description of the electrolyte which is represented by continuous density fields. This framework provided the basis for their subsequent development of a constant-inner-potential formulation, which addresses the more specific question of how the experimentally controlled electrode potential should be represented within a finite DFT calculation~\cite{P6460}.
Grand canonical simulations with varying number of atoms and explicitly varying number of ions are much more challenging, as they typically requires grand-canonical Monte Carlo (GCMC) techniques, and have been applied only for classical force-field simulations.

In 2018, Surendralal et al. introduced the neon electrode~\cite{P5586} as counter electrode to simulate a biased, charged electrochemical interface under periodic boundary conditions. It consists of a monolayer of neon atoms with variable nuclear charge confining the electrolyte region opposite to the slab. Because noble-gas neon atoms have a very large band gap and are thus chemically inert, they do not chemically interact with the system, and by fractional charging or discharging of the neon pseudo-atoms, electrons can be transfer between the metal slab and the neon layer to maintain overall neutrality. The method has been applied to study the initial corrosion at the water-Mg interface.\\
A limitation of the neon electrode approach is the formation of a hydrophobic gap between the counter electrode and the electrolyte, leading to a potential drop within this region~\cite{P7351}. In follow-up work, Deißenbeck et al. in 2021 introduced a canonical thermopotentiostat framework for molecular dynamics simulations at constant temperature and constant applied electric potential~\cite{P6459}. In analogy to theromostats, this method introduces a potentiostat using additional dynamical variables associated with the electrostatic potential, enabling direct sampling of the $NVT\Phi$ ensemble. In 2023, as a next step, the fully solvated electrode has been introduced~\cite{P7364}, which employs a Gaussian counter charge in the electrolyte region to control the charge of the slab. The water density is set by a confining potential between the electrolyte and the periodic image of the electrode generating a vacuum region, in which a dipole correction is applied to eliminate interactions between the periodic images of the system. For a detailed review of these approaches we refer to Ref.~\cite{P7351}.

\subsubsection{Finite field approaches}

Developed within the framework of modern theory of polarization, finite-field methods provide a convenient way to introduce and control macroscopic electric fields in periodic first-principles calculations. Depending on the chosen formulation, the applied electrostatic conditions can be specified in terms of the macroscopic electric-field~\cite{P7395}, polarisation~\cite{P7396}, or the electric displacement field~\cite{P7398}. This is achieved by introducing an extended Hamiltonian that includes the coupling between the electronic system and the applied macroscopic field. 

Finite field methods have recently been applied to the study of electrical double layers at solid–liquid interfaces, including both charged~\cite{P7399} and polar~\cite{P7400} insulator–electrolyte interfaces. The extension of this framework to electrified metallic electrodes immersed in electrolyte solutions has recently been realised~\cite{P7401}. Here, the free charge on the metallic electrode is controlled through the electric displacement field, which determines the macroscopic polarization and electrostatic boundary conditions of the system. At the same time, the simulations explicitly capture the contribution arising from ions in the electrolyte, including specifically adsorbed species at the electrode surface. As the ionic positions are propagated dynamically, the resulting charge distribution and interfacial structure can evolve self-consistently during the simulation. 

\subsection{Classical approaches}

Classical simulations describe electrochemical interfaces using force fields that capture bonded and non-bonded interactions without explicitly treating the electrons in the system. Molecular dynamics and Monte Carlo methods efficiently sample configurational space, typically within the canonical ($NVT$) or the isothermal–isobaric ($NPT$) ensemble, while grand canonical simulations ($\mu VT$) allow ion exchange between the interfacial electrolyte and a bulk reservoir phase, albeit with practical limitations for dense electrolytes. The main limitation, however, is that they rely on fixed atomic species and connectivity, which prevents their application to processes involving chemical reactions or charge transfer. Reactive force fields such as ReaxFF offer a possible route around this restriction~\cite{P0928}. In recent work ReaxFF has been extended to electrochemical interfaces by assigning a free charge density to the electrode surface, compensated by a counterions charge in the electrolyte~\cite{P7523}.

In force field-based approaches, constant charge methods assign fixed charges to electrode atoms, offering computational efficiency but neglecting image charge effects that are essential for accurately describing metallic electrodes. To incorporate electrode polarisation, the GolP force field represents induced charges using permanent dipoles on metal atoms~\cite{P7409,p7410}. This allows to capture image charge effects but with limited transferability and a dependence on adjustable parameters. More recently, core–shell models have enabled a more accurate description of induced polarisation, reproducing image potentials and interfacial properties in good agreement with experiment~\cite{P7411}. These models have been successfully applied to complex metallic interfaces~\cite{P7412}, electrochemical processes~\cite{P7489,P7490}, and, more recently, graphitic materials~\cite{P7413}.

Other constant potential methods maintain a uniform electrode potential by dynamically adjusting atomic charges, thereby providing a more realistic representation of polarisation, although at the expense of greater computational cost. A prominent implementation of the constant potential method  was proposed by Siepmann and Sprik as early as 1995~\cite{P2889}, wherein variable Gaussian charges on the electrode are dynamically adjusted to maintain a uniform electrostatic potential. Several constant potential approaches are now integrated into widely used open-source molecular dynamics packages, including MetalWalls~\cite{P6456} and  ELECTRODE~\cite{P6452} in LAMMPS~\cite{P4473}, enabling, e.g., simulations of complex porous electrodes, such as carbide-derived carbons, without relying on simplified idealised geometries~\cite{P7414,P7423}. Beyond the perfect conductor approximation, the Thomas–Fermi model was subsequently used to account for the screening of the electric field at the interface and describe imperfect conductors such as graphite~\cite{P7486}. 

A detailed comparison of classical simulations performed with and without constant potential across a range of electrode and electrolyte systems has been performed by Zeng et al.~\cite{P7415}. Recently, the constant potential framework has been generalised to multicomponent electrodes by independently controlling the electrochemical potential of each constituent species (CPM$\chi$)~\cite{P7424}. Another extension enables the element-specific parametrisation of charge response and electronegativity, informed by Bader charge analysis, thereby improving the description of chemically heterogeneous electrode materials~\cite{P7425}.

We would like to end this Section with a couple of general remarks regarding the comparison to experiments. 
In experiments, the measured voltage reflects the difference in electron electrochemical potential between the working electrode and the reference electrode. In contrast, atomistic calculations typically do not include an explicit reference electrode. Instead, they provide quantities such as the metal Fermi level or the electrostatic potential of the simulated system. To compare calculated potentials with experimental measurements, these quantities must therefore be aligned to a common reference scale, such as the standard hydrogen electrode (SHE).
A widely used approach is work-function alignment, using Trasatti’s experimentally derived absolute SHE potential of \(4.44 \pm 0.02\ \mathrm{V}\) as a reference.
Alternatively, computational standard hydrogen electrode methods use the mean electrostatic potential in a region of bulk-like solution as its internal reference. A thermodynamic cycle determines the energy offset between this reference and the experimental SHE, thereby mapping the metal Fermi level onto the SHE scale. The hydronium solvation free energy has been used as reference to align free energies to the SHE.

\section{Electrochemistry with MLPs} \label{sec:elecMLP}
\subsection{Overview}

The application of MLPs to electrified interfaces requires several conceptual extensions, and in recent years different methods have been proposed. Local second-generation MLPs do not contain electrostatic interactions preventing the interaction of atoms with electric fields. Third-generation MLPs, which include charges depending on the local atomic environments, in principle allow such interactions. However, charges and multipoles in 3G models cannot adjust to the direction and strength of the field such that these potentials are lacking the ability to describe the field-induced polarization of the system. Moreover, charges depending on the chemical environment only in many cases are unable to represent different oxidation states, unless there are unique differences in the geometric environments, as e.g. in the coordination of Mn$^{3+}$ (Jahn-Teller distorted octahedra) and Mn$^{4+}$ (ideal octahedra) by oxygen atoms in the intercalation compound LiMn$_2$O$_4$~\cite{P5866}. The distinction of different oxidation states by MLPs is also possible for processes like corrosion of metal slabs or the deposition of ions at surfaces, since these processes are usually connected to clear changes in the local atomic geometries. More critical are transition metal ions in solution, since solvation spheres, e.g. in liquid water, can rapidly adjust to different oxidation states and do not represent a structurally stable reference environment. A distinction is thus impossible unless only ions in a consistent single oxidation state are present to avoid contradictory data in the training set. This limitation can in principle be overcome by fourth-generation MLPs, which allow to distribute the charge in the system based on the global chemical composition. As has been demonstrated for the case of Fe$^{2+}$/Fe$^{3+}$, 4G-HDNNPs allow to identify the correct oxidation state depending on the counter ions present in the system, even if they are located outside the local atomic environments~\cite{P6302}. 

\begin{figure*}
  \centering
    \includegraphics [width=0.95\textwidth]{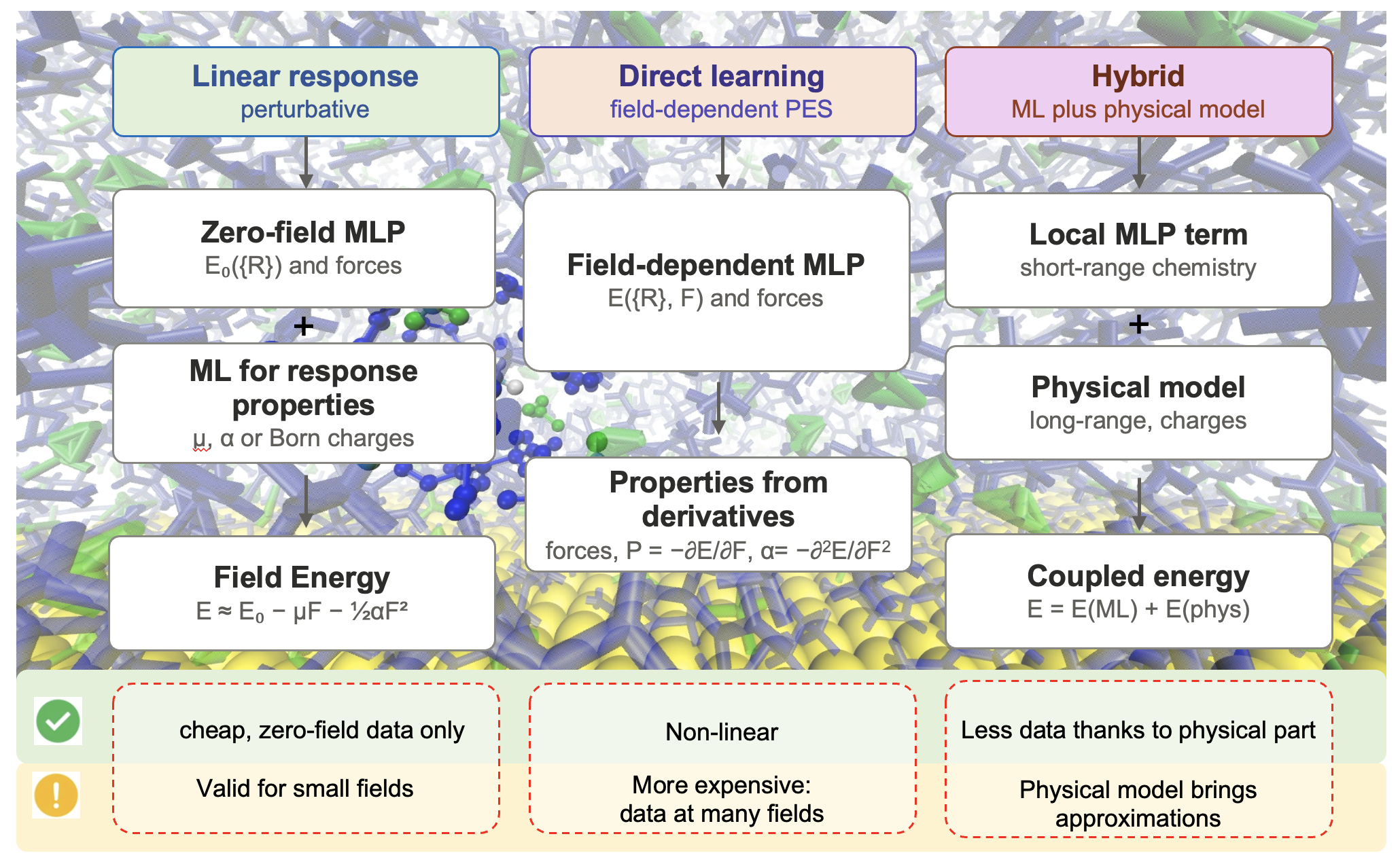}
    \caption{MLP approaches for electrified interfaces. Three classes can be defined according to how the response to an applied electric field is incorporated: (i)  perturbative linear-response models, (ii) directly learning the field-dependent energy landscape, (iii) hybrid approaches coupling MLPs with explicit electrostatic models.}
    \label{fig:MLPelectrochem}
\end{figure*}

Another condition for studying electrochemical interfaces is that the underlying DFT reference data is sufficiently accurate to distinguish different oxidation states, which often requires the use of hybrid functionals or GGA+U calculations for transition metal ions coexisting in different states. 
To address all these limitations and to account for the effect of applied electric fields in MLPs, different strategies have been developed.
These approaches can be broadly classified according to how the response to the applied filed is incorporated: through perturbative linear-response models, by directly learning the field-dependent energy landscape, or by an hybrid approach coupling MLPs with explicit electrostatic models (s. Fig~\ref{fig:MLPelectrochem}).

\subsection{Free-energy calculations and redox potentials}

Computing redox potentials with MLPs is a natural first step toward addressing the more complex problem of electrified interfaces, as it already poses some of the challenges, which we will discuss in the following paragraphs. A redox potential is a free-energy difference: the reversible work of removing an electron from a solute in solution, referenced to a standard electrode. An accuracy of about 0.1 eV (roughly 2 kcal/mol) is typically needed for useful predictions \cite{P7520}, which is difficult to achieve even when using DFT methods directly. The accuracy of DFT thus also forms a boundary for the quality that can be reached by MLPs built from DFT data. 

The free energy differences of redox pairs are typically computed by methods like thermodynamic integration or free energy perturbation along a coupling parameter that connects the oxidized and reduced potential energy surfaces~\cite{P6562}.
The solute and its solvation shell must be sampled thoroughly, since solvent reorganization contributes a significant part of the free energy. Further, the electronic structure must be described accurately, because the delocalization error~\cite{P1990} in semilocal DFT functionals can bias redox energetics substantially, so hybrid functionals or beyond-DFT methods are often required. 

The central difficulty at electrified interfaces is that charge transfer between the electrode and adsorbed or solvated species changes their charge states as the reaction proceeds, whereas commonly used short-range second-generation MLPs do not consider charge, electrostatics and coupling to the electric field. Therefore charge-aware architectures need to be used to treat charged solutes in solution. However, even third-generation MLPs, which include charges and electrostatics explicitly, are unable to simultaneously describe different oxidation states of systems like transition metal ions in solution. Still, employing 2G- and 3G-MLPs is possible if two separate MLPs are constructed to describe different oxidation states separately~\cite{P6336} using training data confined to one of the two states. This approach avoids contradictory information in the training process that would be present when combining ions in different oxidation states but similar geometric environments.  

Recent studies have demonstrated that such MLPs can substantially accelerate first-principles calculations of redox potentials by enabling the extensive finite-temperature sampling required for statistically converged free energies~\cite{P6336,P7511,P7007,P7474}.
Jinnouchi et al. combined MLPs with thermodynamic integration and $\Delta$-learning to obtain redox potentials of aqueous transition-metal ions~\cite{P7007}. To distinguish the oxidized and reduced forms of Fe, Cu and Ag ions, two separate MLPs have been trained for each ion and connected by thermodynamic integration. This strategy was subsequently applied to determine the absolute standard hydrogen electrode potential and a broader set of redox potentials~\cite{P6336,P7511,P7474}. Although the underlying DFT reference calculations include the full electrostatic interactions, the ML representation itself relied on a local second-generation potential. 

In 2024, Wang et al.~\cite{P7511} constructed MLPs that are able to simultaneously describe the reduced and oxidized states of organic molecular redox pairs, which is possible in this particular case since both forms have different molecular structures and stoichiometry enabling local MLPs like the employed DeepPot-SE to distinguish both oxidation states. Very recently, Chen et al. used similar organic molecules to benchmark the performance of several foundation models. They found a good description of proton coupled electron transfer, while mixed results were obtained for electron transfer processes revealing some remaining weaknesses~\cite{P7519}.

\subsection{Linear response models for electrified interfaces}

Moving from homogeneous systems to electrified interfaces adds a further layer of complexity. As the electrode potential becomes an explicit thermodynamic variable, MLPs do not only need to represent the atomic interactions but also the response of the system to an applied potential. A possible solution to this problem is based on linear response theory, which provides an approximate framework for describing how materials respond to externally applied fields. These approaches typically target response properties such as polarisation, polarisability, and Born effective charges~\cite{C0001}. The response function may either be predicted by a dedicated auxiliary model or learned jointly within the same architecture that describes the field-free energy landscape.

An early response-based approach was introduced in 2022 by Shao et al.~\cite{P6913}, who developed a ML scheme for finite-field coupling based on a learned charge response kernel (CRK). Here, an atom-condensed CRK is learned from molecular polarizability and used to predict the redistribution of atomic charges in response to an applied electric field. 
Training and prediction of the CRK are performed with PiNet, a graph convolution neural network (GCNN) implemented in PiNN~\cite{P5719}.
This approach provides a linear-response description of electronic polarization, with the response charge obtained from the learned charge–field susceptibility, suggesting a route toward incorporating finite-field effects into atomistic simulations without explicitly training separate potentials for different fields. This can be done in a hybrid framework, PiNNwall~\cite{P7451}, which will be described in more detail below in the section dedicated to hybrid models.
PiNN has been recently extended~\cite{P6910} to include an equivariant model for energies and forces (PiNet2) and dedicated neural networks for dipoles and atomic charges, polarizability, and the CRK. 

In 2023, Schienbein introduced an E(3)-equivariant graph neural network to predict Born effective charges, referred to as the atomic polar tensor (APT)~\cite{P6484}. Building on this work, in
2024 Joll et al.~\cite{P7402} devised the perturbed neural network potential (PNNP) to describe the influence of external electric fields on condensed-phase systems. In this framework, the total force acting on each atom is calculated by combining the intrinsic forces of the unperturbed system represented by a committee of 2G-HDNNPs (c-HDNNPs) with a field-induced contribution proportional to the field through the APT. 
Both components, c-HDNNPs and APT are trained solely on data obtained in the absence of an external electric field. Applying this framework, it has been demonstrated that electric fields can induce a transition from vehicular to structural ionic transport in aqueous alkali-metal solutions, revealing how field-dependent changes in ion hydration and ligand exchange govern microscopic charge transport~\cite{P7322}.

A similar example of augmenting MLPs with learned electrical response quantities for simulations under external fields is the approach by Shimizu et al.~\cite{P7469}, who developed a neural network model to predict configuration-dependent Born effective charges. The predicted Born effective charges were combined with a conventional neural-network potential to evaluate the forces induced by an external electric field, enabling MDs simulations of field-driven Li migration in crystalline and amorphous Li$_3$PO$_4$. The authors showed that the off-diagonal components of the Born effective-charge tensor can contribute to the transverse forces acting on migrating Li ions.  
Later, in 2025 Kutana et al.~\cite{P7471} expanded on this by developing an equivariant graph convolutional neural network that directly predicts full tensorial Born effective charges and with applications to Li$_3$PO$_4$, perovskite oxides, and ZrO$_2$.

In 2025 Falletta et al. ~\cite{P7403} develop a unified, differentiable ML framework in which the electric enthalpy is learned as a function of both the atomic configuration and the applied electric field. Here polarization, Born effective charges, and polarizability are obtained as derivatives of the same learned electric enthalpy, ensuring the appropriate physical symmetries, conservation laws, and relations between response properties. The model is trained using DFT data for energy/enthalpy and forces at zero-field. For finite-field simulations, the zero-field Born charges and polarizability are then used to add the field-induced contributions to the forces and polarization. The authors demonstrate the approach for $\alpha$-quartz and BaTiO$_3$, including dielectric properties and large-scale simulations of ferroelectric hysteresis.
Moreover, it has been proposed to train a MLP on DFT energies and forces while introducing environment-dependent latent atomic charges, from which polarization and Born effective charges can be obtained~\cite{P7262}. These are then used to couple the learned potential to an external electric field and simulate the resulting electrical response. Another method to determine Born effective charges using invariant descriptors has been presented by Kresse and coworkers in 2026~\cite{P7521}.

In 2025 Bergmann et al.~\cite{P7255} introduced RAZOR, a ML framework for electrified interfaces that learns first-order electrochemical response parameters, including the work function and Born effective charges. By expanding energies and forces as a function of excess charge, RAZOR enables bias-dependent molecular dynamics with charge-responsive forces. Applied to Cu(100)/OH interfaces, it achieved near-DFT accuracy while strongly reducing the computational cost of the simulations.

The relation between charges trained to represent long-range electrostatic interactions as contribution to the PES and dynamic Born effective charges for linear response has been studied by St\"ark et al.~\cite{P7260} with the recommendation to treat both properties independently, since only Born effective charges can be rigorously determined from electronic structure calculations, while atomic point charges are model-dependent and serve the purpose of constructing the PES itself.

\subsection{Direct learning of the field-dependent energy surface}

Direct ``end-to-end'' learning approaches offer the greatest flexibility. Here, the full potential-dependent energy surface is learned without relying on response theory, which allows to naturally capture non-linear field effects. This, however, requires extensive DFT data spanning the relevant range of potentials or charge states, and their extrapolation beyond the training domain is often less reliable than response theory-based methods. 

It should be noted that in principle many standard MLPs can be used directly to represent a particular charge state of an electrode without modification if the corresponding DFT training data is available. However, in this case the resulting MLP is unlikely to be transferable to systems with different chemical composition, preventing upscaling as a major advantage of MLPs. Consequently, a more promising approach is to construct models, in which the applied potential, electric field, or electron number is supplied explicitly as an additional input to make MLPs more broadly applicable to a range of field strengths, directions, or charge states within a single model. We will report here a few examples, which adopt this end-to-end strategy. 

Early work incorporated external electric fields explicitly into MLPs for molecules. In 2021 FieldSchNet~\cite{P6108} extended the message passing SchNet architecture to include the interaction of a molecule with its surroundings through a set of vectorial features describing the system. Although FieldSchNet aimed at isolated molecules in solution rather than electrified interfaces, it established the idea of feeding field information into the network itself, and it paved the way for later periodic and interfacial models.

In 2023, Zhang and Jiang~\cite{P7404} introduced the field-induced recursively embedded atom neural network (FIREANN), which incorporates the external electric field directly into the atomic environment descriptor through a field-induced embedded atom density (FI-EAD). Response properties, including dipole moments and polarisabilities, are obtained analytically as derivatives of the predicted energy with respect to the electric field and can be learned jointly during training. By training only on atomic forces, FIREANN avoids the multivalued polarisation problem in periodic systems while accurately reproducing dipole moments and infrared spectra for liquid water. An extension of FIREANN was subsequently proposed~\cite{P7277} by coupling it with a ML electron density response  model~\cite{P7405}, enabling the prediction of both field-dependent atomic forces and charge redistribution from the first-principles data. Applied to the Au(100)/water interface, the framework accurately reproduced interfacial electrostatic properties, including the Helmholtz capacitance, while achieving approximately four orders of magnitude greater computational efficiency than ab initio MD.

In 2023, Chen et al.~\cite{P7291} developed a dual-learning framework for electronically grand-canonical simulations, combining two neural networks based on Behler–Parrinello atom-centered symmetry functions. One network predicts atomic charges from atomic configurations and electrode potential, while the other learns environment-dependent electronegativities. The potential-dependent descriptors introduce spatially decaying electric-field effects near the interface, enabling a physics-informed charge–electronegativity energy formulation. Applied to hydrogen evolution on Au(111), the model achieved high accuracy, transferred across different metals, and substantially accelerated barrier calculations compared with DFT.

Zhou et al.~\cite{P7293} in 2025 introduced the explicit electric potential machine-learning force field (EEP-MLFF), which incorporates the applied electrode potential directly into the atomic descriptor to learn the potential-dependent energy surface. Applied to Cu cluster formation on MoS$_2$, the model achieved good agreement with DFT, although it was trained over a relatively narrow potential window (-0.5 to +0.6 V vs. SHE).

Further advanced grand-canonical equivariant machine-learning potentials for electrochemical interfaces have been developed by Chen et al.~\cite{P7283}, in which a global excess charge is introduced as an additional variable and its redistribution among atoms is learned through a multi-head attention mechanism, thereby capturing the nonlocal redistribution of excess charge across the interface. The model directly learns the potential-dependent energetics and charge distribution from grand-canonical DFT data, enabling simulations of electrochemical reactions under constant-potential conditions.

Furthermore, in 2025, Sun et al. introduced DP-N$_e$, a modified Deep Potential model~\cite{P7292} that incorporates the total electron number (N$_e$) as an additional input to the embedded representation. The model simultaneously predicts the work function, (W=-$\partial$E(R,N$_e$)/$\partial$N$_e$), enabling to capture the dependence of the electronic chemical potential on both atomic configuration and charge state. Coupled with a grand-canonical path-integral hybrid Monte Carlo (GC-PIHMC) framework, DP-N$_e$ was used to investigate nuclear quantum effects at electrochemical interfaces, revealing that proton tunneling lowers the activation barrier for the hydrogen evolution reaction on Pt(111) by 0.09–0.13 eV and increases reaction rates by up to two orders of magnitude at room temperature.

Building on the MACE framework, Wang et al.~\cite{P7279} developed a constant-potential machine-learning force field (CP-MLFF) that extends MACE to grand-canonical simulations of electrochemical interfaces. By incorporating the total electron number as either an atomic or graph-level feature, the model learns energies, forces, and the Fermi level under constant-potential conditions. Its capability was demonstrated for CO$_2$ reduction on a Ni–N–C catalyst. 
Although incorporating the total electron number allows to describe grand-canonical charging, the use of a global, size-dependent descriptor might affect transferability across system sizes.

Very recently, Chen et al.~\cite{P7452} developed a constant-potential reactor framework that combines a variable-charge neural network potential (veNNP) with constant-potential molecular dynamics and enhanced-sampling active learning. The veNNP explicitly accounts for variations in the system charge and simultaneously predicts energies, forces, atomic charges, and the corrected Fermi level. During the MD run a potentiostat based on a modified Nos\'e-Hoover Lagrangian formalism  enables charge exchange with an electron reservoir to maintain a prescribed electrode potential. This framework is coupled with an explicit–implicit hybrid solvent description and has been applied to study the role of alkali-metal cations on CO$_2$ reduction at Au–water interfaces.

Finally, Bianchi et al.\cite{P7299} addressed the high computational cost of training MLPs for constant-potential electrochemical interfaces by introducing TRECI, a data-efficient transfer-learning workflow for electronically grand-canonical molecular dynamics. The reference configurations are generated under constant-potential conditions, in which the electrode charge is allowed to fluctuate to maintain a prescribed potential. Instead of training independent models from scratch for each potential, TRECI transfers pretrained atomistic representations and employs shared descriptors with potential-specific regression heads, while active learning selectively adds new reference configurations. This approach was applied to the Cu(111)–water interfaces, showing that constant-potential molecular dynamics over a broad potential range was possible using approximately 1000 configurations per bias. This enabled the use of quite expensive meta-GGA reference calculations and revealed potential-dependent restructuring of interfacial water. 

\subsection{Hybrid models for electrified interfaces}

Rather than capturing the full dielectric response of metal/electrolyte interfaces within a single ML model, hybrid approaches partition the system into physically distinct components, each described using the representation most appropriate for its underlying physics. For example, the molecular structure and interactions within the electrolyte can be described using an atomistic MLP, while the metallic electrode can be represented by another dedicated model covering charge redistribution and electronic response.  

Building on the CRK approach described above~\cite{P6913}, Dufils et al. introduced PiNNwall~\cite{P7451}, a framework that couples ML models of atomic interactions and charge response with the classical electrostatic simulation package MetalWalls~\cite{P6456} to model heterogeneous electrified electrode-electrolyte interfaces. The ML model provides the short-range atomic interactions and charge response, while MetalWalls describes the long-range electrostatics and the response of the electrode to an applied potential. PiNNwall enables finite-field simulations of heterogeneous interfaces while retaining the computational efficiency of MLPs and the explicit electrostatic treatment required for electrochemical systems. 

A related hybrid strategy was also introduced by Hu et al.~\cite{P7497}, who developed DP-QEq, by coupling a Deep Potential model for short-range interactions with a charge-equilibration description of the long-range electrostatic response. Conceptually, this builds on the same general idea of combining a learned short-range potential with an explicit fluctuating-charge electrostatic model as in PiNNwall, although DP-QEq uses a charge equilibration formulation~\cite{P1448} in place of the Siepmann–Sprik electrode model~\cite{P2889}. The resulting framework enables constant-potential simulations of large electrochemical systems and was applied to Li-metal/electrolyte interfaces to describe Li deposition and the early stages of dendrite formation. 

Zhu and Cheng developed a hybrid framework of MLPs, ec-MLP~\cite{P7123,P7297}, combining a local-descriptor-based Deep Wannier model~\cite{P6260} to describe the dielectric response in the electrolyte and a Siepmann–Sprik model~\cite{P2889} to treat nonlocal charge transfer in metal electrodes. Benchmark calculations have demonstrated that this hybrid framework accurately reproduces both the potential-dependent structure of interfacial water and the characteristic bell-shaped differential capacitance of Pt(111)/electrolyte interfaces. This framework has a modular design where both the electrode and electrolyte models can be readily replaced, as improved MLPs become available. 

Along similar lines, Cheng and coworkers introduced a hybrid approach~\cite{P7421}, which combines Latent Ewald Summation to determines charges in the electrolyte and the Siepmann-Sprik model to obtain electrode charges minimizing the electrostatic energy under an external bias.
Using this approach, they reproduce the chemisorbed and physisorbed water layers at Pt(111) which respond asymmetrically to applied bias. They also show that K$^{+}$ and F$^{-}$ ions instead accumulate primarily within the compact layer without substantially perturbing the local water structure, but effectively screening the applied electric field. The simulations further reveal atomistic proton-transfer mechanisms that connect chemisorbed, physisorbed and bulk water through Grotthuss-like transport~\cite{P7421}. 

In general, the modular architecture of hybrid models enables systematic improvements to the electrode model. For example, in the charge equilibration of the electrode in the Siepmann-Sprik approach the fixed width Gaussian charges can be replaced by variable width charges to describe local heterogeneity, as e.g. in alloy electrodes, more accurately, while providing a rigorous treatment of the Fermi level and electrode potential. 
For applications requiring accurate electronic spatial distribution, DFT-quality charge density models, such as the symmetry adapted learning of three dimensional electron densities (SALTED)~\cite{P7313} can be readily incorporated~\cite{P7312}. SALTED has also been integrated into hybrid frameworks alongside classical electrolyte force fields.  In a study of the Au(100)/NaCl interface~\cite{P7311}, it was used to provide a QM/MM-level description by learning the gold metal electron density, while the electrolyte was treated classically. 

A key limitation of hybrid frameworks is the partitioning of the system into separate dielectric regions, which raises questions about its ability to capture interfacial electronic effects consistently. Chemisorption may appear challenging to describe if the electrolyte is represented by fixed-charge species. However, this limitation can be overcome by describing charge transfer through interfacial electronic polarisation, where electron redistribution within the adsorbate under constant-potential conditions provides an equivalent representation of chemisorption~\cite{P7297}.
Further developments could also allow for a dynamical definition of the partitioning of the separate dielectric regions, in the spirit of the adaptive QM/MM models~\cite{P4998}, in response to the instantaneous atomic configuration. In the present context, such an approach could allow the regions described by different dielectric models to adapt to the local environment, for example as molecules approach, leave, or reorganize at the electrode interface.

An additional challenge is that reproducing the local charge distribution does not necessarily ensure an accurate description of the macroscopic electrostatic field. Recent work by Yang et al.~\cite{P7296} showed that even highly accurate local ML predictions can fail to reproduce the weak macroscopic fields at electrochemical interfaces, due to the much larger thermal fluctuations of the liquid phase. Their macro-dipole-constrained learning scheme addresses this issue by incorporating the total dipole moment of the simulation cell into the learning procedure, thereby enforcing consistency between the learned local charge distribution and the global electrostatic response. 

More generally, hybrid strategies can also combine atomistic energy models with dedicated constant-potential algorithms, separating the description of the atomic interactions from the control of the electronic chemical potential. In this direction, Zhang et al.~\cite{P7516} developed a flexible constant-potential MD framework within the i-PI environment~\cite{P7538} that is independent from any specific electronic structure code. The framework introduces an explicit electronic degree of freedom and a dedicated potentiostat to regulate the electronic chemical potential. A mixed-Hamiltonian interpolation scheme over two adjacent integer-charge states provides an effective fractional-electron description without requiring modifications to the underlying electronic structure code. The method was tested for an Al(111) surface and subsequently combined with enhanced sampling to investigate potential-dependent CO$_2$ reduction on NiN$_4$-doped graphene~\cite{P7516}. 

\section{Discussion and Perspectives}\label{sec:discussion}

State-of-the-art MLPs now permit to address some of the long-standing challenges associated with the atomistic modelling of electrochemical interfaces. One of the most immediate advantages is the possibility of extending simulation timescales from the picosecond regime towards nanoseconds and beyond, thereby improving statistical sampling and enabling the exploration of processes that are difficult to capture using computationally demanding ab initio MD, such as the electric double layer equilibration~\cite{P7346}. When combined with enhanced-sampling techniques, MLPs can bring simulations closer to the regimes relevant to experiments, addressing, e.g., interfacial restructuring, solvent reorganization, ion transport and  chemical reactions. Equally important is the ability to simulate much larger systems, allowing for more complex and realistic representations of electrode surfaces, electrolytes and their interfaces. Together, these advances offer a promising route towards bridging the gap between atomistic simulations and experiment.

These opportunities, however, should not distract from the fact that ML-based approaches inherit many of the challenges of the underlying electronic-structure methods. Most notable, even the question how to best perform electronic structure calculations of electrified interfaces at constant potential is not yet fully settled~\cite{P7351}. Moreover, beyond this specific challenge of electrified interfaces,
the reliability of MLPs in general depends on the quality, accuracy and configurational coverage of the reference data on which they are trained. MLPs are by far most commonly trained on DFT data and thus suffer from the intrinsic limitations of a particular exchange–correlation functional, including uncertainties in the description of adsorption energetics, charge transfer, reaction barriers and aqueous electrolyte properties~\cite{P6212}.

Not only the quality but also the amount and distribution of training data is important, as the transferability of MLPs is often limited when applied to chemical environments or electrochemical conditions that are insufficiently represented in the training set.  Progress therefore depends on both, developing more efficient learning algorithms achieving better transferability while requiring less data, but also on 
the use of more accurate reference methods for generating this training data. Interestingly, recent rather complex ML models relying on message passing~\cite{P5368} and equivariant features~\cite{P6017} seem to take a step in this direction in spite of being at odds with traditional wisdom that more parameters should require more training data.

To improve the accuracy, in particular for electrolyte subsystems and liquid water, MLPs based on coupled cluster data are now emerging~\cite{P6360,P6362,P6321,P7525}.
Regarding the description of electrodes, challenges arise from the increasing interest in strongly correlated materials like transition metal oxides. In these systems, the electronic structure often is not adequately captured by GGA or even hybrid functionals requiring advanced electronic-structure approaches as well as careful validation against experimental observables. Currently, the choice of the reference method for constructing MLPs necessarily remains a compromise to achieve an overall reasonable and balanced description of all parts of a system.

Another aspect worth noting is that the role of nuclear quantum effects (NQE) in electrochemical processes is still largely unexplored, although a significant impact is well-known for proton transfer events, which are central in electrochemistry. This is mainly due to 
the high computational costs of explicitly including NQEs. MLPs will substantially contribute to making such simulations more accessible in the near future.

In parallel, a promising direction 
is the development of ML-based electronic-structure methods that aim to provide higher levels of quantum-chemical accuracy at a much reduced computational cost. Recent advances in machine-learned exchange–correlation functionals, such as the neural network-based functional Skala designed to bridge the gap between high-accuracy wavefunction methods at the computational efficiency of semi-local DFT~\cite{P7528}, and also attempts to accelerate electronic structure calculations through learned Hamiltonias~\cite{P7417} illustrate this possibility. 

Next to the level of electronic structure theory, further possible uncertainties arise from the very limited size of the reference structures included in the training sets. In principle, the construction of MLPs for large-scale simulations from much smaller training systems is a fundamental reason for the success of modern MLPs~\cite{P5537,P6065}. Still, in case of electrified interfaces finite-size effects may influence the mean electrode potential and its fluctuations by constraining interactions between adsorbates, water molecules and ions. Moreover, surface defects, e.g. in oxides, can change the system's Fermi level leading to long-range or even global changes in the atomic interactions that are hard to include in small model systems. Therefore, in such cases analyzing the physical information included in the reference data is important. 

Beyond fully atomistic simulations, ML is also contributing to advance coarse-grained modelling, in which selected molecular degrees of freedom are represented through effective interactions. Such approaches are particularly attractive for complex electrolytes, including ionic liquids~\cite{P7529} and polymer electrolytes~\cite{P7530}, where the more complex molecular structure and slow dynamics may require length and timescales beyond those accessible to fully atomistic simulations. A complementary avenue is the integration of ML with classical density functional theory, which provides a statistical-mechanical framework for describing the inhomogeneous structure of liquids~\cite{P7531} and electrolytes near (electrochemical) interfaces~\cite{P7532}. These works are part of a broad effort to develop ML methods for representing or improving free-energy functionals~\cite{P7533}.
Ultimately, the simulation of electrochemical processes will continue to rely on multiscale approaches that integrate models with different levels of resolution, combining their complementary strengths within a unified framework.

A fully grand-canonical description, not only of the electrons but also, consistently, of the ions, is highly desirable, yet has proved challenging to implement in ab initio molecular dynamics.
In the context of force-field-based simulations, however, various approaches to simulate open systems at constant chemical potential $\mu$ (C$\mu$MD) have been developed, which allow molecules and ions to exchange between a finite simulated region and an external reservoir~\cite{P7534}. A particular case of C$\mu$MD is constant-pH MD, in which the protonation states of reactive sites can be adjusted dynamically. Such methods are widely used in biomolecular simulations~\cite{P7535} and, to a lesser extent, have also been applied to surfaces, e.g. to oxide–water interfaces to investigate protonation fluctuations and surface electrostatic potentials at goethite nanoparticles~\cite{P7536}. Both C$\mu$MD and constant-pH methods are particularly relevant to electrochemical interfaces, where protonation equilibria can strongly influence interfacial structure and reactivity, while electrode growth and dissolution require simulations at constant chemical potential. These approaches could be extended to MLPs, combining the flexibility of constant-$\mu$, constant pH simulations with the accuracy of electronic-structure methods.

A further unresolved challenge in the atomistic modelling of electrochemical interfaces is the description of lateral charge transport within the electrode. Although constant-potential methods, for example those employing potentiostats, allow charge exchange normal to the electrode surface, charge redistribution parallel to the surface is assumed to occur instantaneously under periodic boundary conditions, effectively imposing a uniform electric field~\cite{P7351}. This assumption neglects the finite electrical conductivity and surface resistance of real electrodes, limiting the ability of current approaches to capture the dynamics of lateral charge equilibration. Advances in MLPs for modelling non-equilibrium dynamics may also help address these challenges.

Apart from all these considerations, there is also a clear need to further advance current state-of-the-art MLPs with respect to both, the explicit inclusion of physical models and the improvement of transferability. Long-range electrostatic interactions are central for the application to electrified interfaces, and as briefly discussed in Section~\ref{sec:MLPs}, many different methods have been proposed to determine the required charges. There are essentially two different groups of methods, using either reference charges and multipoles from electronic structure calculations~\cite{P2041,P2962}, which are no physical observables and numerically depend on the specific partitioning scheme employed, or latent charges determined to reproduce the energies and forces in the reference data~\cite{P7233}, which are more difficult to interpret. Beyond charges and multipoles, recently a lot of effort has been put into learning charge densities~\cite{P7313}, which in the future will result in more refined methods.

The practical usability of MLPs has substantially improved in the past few years with a large variety of open-source software packages for training and applying MLPs now being available. In many cases, simulations are greatly facilitated by direct interfaces to widely distributed software frameworks like LAMMPS~\cite{P4473}, i-PI~\cite{P7538} and ASE~\cite{P5915}, where other types of potentials can in many cases be replaced by MLPs in a straightforward way. This provides immediate access to many types of simulations beyond standard MD, like enhanced sampling and free energy methods. Some extensions in workflows may be required, however, in case of linear response methods, in which additional properties like Born effective charges need to be predicted on-the-fly. Still, although tempting, the replacement of conventional force fields or electronic structure methods by MLPs has to be done with care, since the physical model of the selected MLP must be suitable for the problem to be studied. For instance, systems with non-local dependencies in the system cannot be captured by local second-generation potentials, and require the choice of more advanced methods~\cite{P5977}, which also come at increased computational costs. Experience and insights into the physics of the system of interest is required for this choice, in full analogy to the wealth of experience that has been built up in the community regarding, e.g., the choice of appropriate exchange correlation functionals in DFT calculations. 

Regarding transferability, rapid progress has been made in recent years in the construction of general purpose foundation models with very broad applicability across the periodic table~\cite{P7067}. They rely on large and constantly increasing databases covering many materials and molecular systems. Still, the structures included in these databases will always remain restricted to a small part of the huge chemical space such that system-specific fine-tuning remains an important step. Further, with increasing applicability of transferable MLPs, validation becomes a central task, as generating stable trajectories is no evidence for correctness. This validation is very challenging, since the confirmation of results by DFT calculations is in most cases prohibitively expensive. Simple accuracy quantifiers like energy and force errors of training and validation sets are not reliable, since they strongly depend on the composition of these datasets and do not cover unseen regions of configuration space. Moreover, there is no clear relation between energy and force errors and the resulting errors of other physical properties. Therefore, any meaningful validation workflow must include a variety of physical observables, which requires insights into the physics of the system and the awareness of the limitations of the underlying reference electronic structure method.
Equally important is the comparison with experimental data, which must itself be critically assessed and, where possible, standardized to ensure consistency across different measurement techniques and conditions. This requires a careful evaluation which observables can be computed reliably and compared meaningfully with experiment, establishing a common ground between simulation and measurement. Robust validation rests on bringing together physical insight, computational accuracy and experimental evidence, while recognizing the uncertainties and limitations inherent in each.

Overall, a lot of progress had been made in extending the applicability of MLPs to the field of electrified interfaces. While molecular dynamics studies of solid-liquid interfaces have been carried out for many years, there are now several approaches, which allow to include the effect of electric fields and electrode potentials directly into the atomic interactions. Some challenges remain to be solved in future developments, but combining MLPs with physical insight and rigorous validation practices will enable the electrochemistry community to advance fundamental knowledge and strongly support future technological advances.

\section{Author contributions}  

Both authors contributed equally to all aspects of this work.

\begin{acknowledgments}
This project has been funded by the Deutsche Forschungsgemeinschaft (DFG, German Research Foundation) under Germany’s Excellence Strategy EXC 2033 RESOLV (project-ID 390677874). 
M.S. acknowledges support by the Research Training Group
“Confinement-Controlled Chemistry” funded by the DFG - GRK2376/331085229. J.B. thanks the DFG for support in collaborative research center CRC 1633 ``Pushing Electrons with Protons'' (project C04, project number 510228793). We thank K. Nikolas Lausch for generating Figure 1.
\end{acknowledgments}

\bibliography{literature}


\end{document}